\documentclass[prb,aps,amssymb,showpacs,twocolumn]{revtex4}
\usepackage{graphicx}

\begin{document}

\title{Mutual Chern-Simons Theory of Spontaneous Vortex Phase}
\author{Xiao-Liang Qi and Zheng-Yu Weng}
\affiliation{Center for Advanced Study, Tsinghua University, Beijing, 100084, China}
\date{\today}

\begin{abstract}
We apply the mutual Chern-Simons effective theory (Phys. Rev. B
\textbf{71}, 235102) of the doped Mott insulator to the study of the
so-called spontaneous vortex phase in the low-temperature pseudogap
region, which is characterized by strong unconventional
superconducting fluctuations. An effective description for the
spontaneous vortex phase is derived from the general mutual
Chern-Simons Lagrangian, based on which the physical properties
including the diamagnetism, spin paramagnetism, magneto-resistance,
and the Nernst coefficient, have been quantitatively calculated. The
phase boundaries of the spontaneous vortex phase which sits between
the onset temperature $T_{v}$ and the superconducting transition
temperature $T_{c}$, are also determined within the same framework.
The results are consistent with the experimental measurements of the
cuprates.
\end{abstract}

\maketitle



\section{Introduction}

Since the 1986 discovery\cite{bednorz1986}, the high-$T_{c}$ cuprate
superconductors have attracted strong interest both theoretically
and experimentally. However, a well-accepted understanding is still
elusive after two decades' efforts. The main difficulty comes from
the strongly correlated nature of the electronic dynamics and the
complex phenomena observed in experiments. Although there have been
a lot of theoretical proposals available in describing some aspects
of the experimental observations, the most important challenge to a
microscopic theory of the cuprates is how to provide a consistent
understanding of the global and universal features of the whole
phase diagram.

The mutual Chern-Simons gauge theory proposed in Ref.
\onlinecite{kou2005} aims at facing this challenge. This theory is
based on the phase string formalism of the $t-J$
model\cite{weng1997,weng1998}, in which the charge and spin degrees
of freedom are both described by bosonic fractionalized
fields---called holon and spinon, respectively. The fermionic
statistics of the electron is taken into account by the mutual
topological interaction between each spinon and holon, mediated by
the mutual Chern-Simons gauge fields. The advantages of this
approach are i) naturally including two ordered phases, \emph{i.e.,}
antiferromagnetic (AF) and superconducting (SC), in the phase
diagram; ii) explicitly incorporating the strong mutual influence
between the charge and spin degrees of freedom via the mutual
Chern-Simons gauge structure. In this theory, the bosonic spinons
form the singlet pairing at a characteristuc temperature $T_{0}$,
whose maximum $\sim J\simeq 1,480$ $\mathrm{K}$ at half-filling
while monotonically decreases with doping. Physically, $T_{0}$
stands for the temperature scale below which the short range AF
correlations start to grow from the length scale of the lattice
constant.\cite{gu2005} The regime at $T<T_{0}$ is called the
\emph{upper pseudo-gap phase} (UPP), which is the ``matrix" of all
the lower temperature phases for the underdoped system, including
the AF and SC ordered states. With each spinon playing the role of a
$\pi $-vortex of the holon field and vice versa, the
superconductivity is described by the holon condensation with the
spinon-pair confinement, and the AF ordered state is depicted by the
spinon condensation with the holon self-localization\cite{kou2005}.
Here the global phase diagram is essentially decided by the mutual
duality of the mutual Chern-Simons theory.

In this paper, we will apply this theory to a low-temperature
phase embedded in the UPP, which is described by the holon
condensate with {\em unconfined} spinons. Physically, such a
regime corresponds to an unconventional SC fluctuational region,
in which strong SC fluctuations are driven by the low-lying spin
excitations. In other words, such a phase is both a spin liquid
and a vortex liquid, named by spontaneous vortex phase (SVP). A
semi-classical mathematical description of such a region in the
language of the generalized Ginzburg-Landau equation has been
given in Refs. \onlinecite{weng2002,weng2006}. The main goal of
the present work is to provide a mathematically self-contained
effective theory, based on the mutual Chern-Simons gauge theory,
which can work beyond the Ginzburg-Landau formalism. Starting from
this effective theory, the physical quantities including the
diamagnetism, spin paramagnetism, magneto-resistance, and the
Nernst effect are studied, which are compared with the
experimental results. The phase boundary of the SVP in the $T-H$
(temperature-magnetic field) plane is decided by the
superconducting transition temperature and characteristic magnetic
field, $T_{c}$ and $H_{m}$, and
the holon condensation onset temperature and critical magnetic field $%
T_{v},H_{c2}$. All these four quantities and their doping dependence
are calculated based on the leading order approximation in this
effective theory, which show qualitative consistency with the
experiments.

The remainder of this paper is organized as follows. In Sec. II,
the effective description for the SVP is derived from the
mutual-Chern-Simons gauge field theory of the phase string model.
Sec. III is contributed to the calculation of physical quantities
at the mean-field approximation level of the effective theory,
which include the diamagnetism, paramagnetism, magneto-resistance
and Nernst coefficient. Then the calculation of the phase boundary
beyond the mean-field approximation is given in Sec. IV. Finally,
the conclusion and discussions are presented in Sec. V.

\section{Mutual Chern-Simons Theory of Spontaneous Vortex Phase}

\subsection{Mutual Chern-Simons Effective Theory for a doped Mott Insulator}

The mutual Chern-Simons theory \cite{kou2005} is a field theory
description of a doped Mott insulator. In such a theory, a mutual
topological interaction between spin and charge degrees of freedom,
due to the phase string effect\cite{weng1997,weng1998} in the $t-J$
model, is simply captured by a mutual-Chern-Simons term. The total
lattice Euclidean Lagrangian is given as
follows\cite{kou2005}%
\begin{widetext}
\begin{eqnarray}
L&=&L_{h}+L_{s}+L_{CS}, \label{LMCS}\\
L_{h} &=&\sum_{I}h_{I}^{\dagger }\left[ \partial
_{0}-iA_{0}^{s}(I)\right] h_{I}-t_{h}\sum_{\left\langle
IJ\right\rangle }\left( h_{I}^{\dagger
}e^{iA_{IJ}^{s}}h_{J}+h.c.\right) +\mu \left(
\sum_{I}h_{I}^{\dagger
}h_{I}-N\delta \right)+\frac {u} 2\sum_I\left(h_I^\dagger h_I\right)^2, \label{Lh} \\
L_{s} &=&\sum_{i\sigma }b_{i\sigma }^{\dagger }\left[ \partial
_{0}-i\sigma A_{0}^{h}(i)\right] b_{i\sigma
}-\frac{J}2\sum_{\left\langle ij\right\rangle \sigma
}\Delta_{ij}^s\left( b_{i\sigma }^{\dagger }e^{i\sigma
A_{ij}^{h}}b_{j-\sigma }^{\dagger }+h.c.\right) +\lambda \left(
\sum_{i\sigma }b_{i\sigma }^{\dagger }b_{i\sigma }-N\left(
1-\delta \right) \right), \label{Ls}\\
L_{CS}&=&\frac{i}{\pi }\sum_{I}\epsilon ^{\mu \nu \lambda }A_{\mu
}^{s}(I)\partial _{\nu }A_{\lambda }^{h}(i),  \label{lcs}
\end{eqnarray}
\end{widetext}
in which $L_{h}$ and $L_{s}$ describe the dynamics of the matter fields ---
\emph{bosonic} spinless holon, $h_{I}$, and \emph{bosonic} neutral spinon, $%
b_{i\sigma }$, respectively. The chemical potential $\lambda$ and
$\mu$ terms are included to enforce the total number constraint of
spinon and holon, respectively. In the $\mu $ term, $\delta $
denotes the doping concentration and $N$ the total number of lattice
sites, and the last term in $L_{h}$ is introduced to account for the
on-site repulsion between the holons, which may be regarded as a
softened hard-core condition.

The two matter fields $h_I$ and $b_{i\sigma}$ minimally couple to two \textrm{U(1)} gauge fields $%
A^{s}$ and $A^{h}$, respectively, in $L_{h}$ and $L_{s}$.
Physically, the mutual-Chern-Simons coupling $L_{CS}$
[Eq.(\ref{lcs})] entangles two otherwise independent gauge fields,
$A^{s}$ and $A^{h}$, to realize the topological constraint due to
the phase string effect. This may be directly understood by
considering the equations of motion for the temporal components
$A_{0}^{h}$ and $A_{0}^{s}$:
\begin{eqnarray}
\frac{\partial L}{\partial A_{0}^{s}(I)} &=&0\Rightarrow \epsilon
^{\alpha \beta }\Delta _{\alpha }A_{\beta }^{h}(i)=\pi n_{I}^{h}
\label{constraint1} \\
\frac{\partial L}{\partial A_{0}^{h}(i)} &=&0\Rightarrow \epsilon
^{\alpha \beta }\Delta _{\alpha }A_{\beta }^{s}(I)=\pi
\sum_{\sigma }\sigma n_{i\sigma }^{b} \label{constraint}
\end{eqnarray}%
In other words, the holon (spinon) number
field $n_{I}^{h}$ ($n_{i\sigma }^{b})$ determines the gauge-field strength of $A^{s}$ ($%
A^{h})$, as if each matter particle (holon or spinon) is attached to
a fictitious $\pi $ flux tube seen by the different species. Note
that in the
lattice version of $L_{CS}$ [Eq.(\ref{lcs})], $\partial_{\alpha}=\Delta_{%
\alpha}$ for the spatial components with $\Delta _{\alpha }A_{\beta }^{h}(i)
\equiv A_{\beta }^{h}(i+\hat{\alpha})-A_{\beta }^{h}(i)$ and $\Delta
_{\alpha }A_{\beta }^{s}(I)\equiv A_{\beta }^{s}(I)-A_{\beta }^{h}(I-\hat{%
\alpha}).$ Here the indices $\alpha $ and $\beta $ will be always used to
denote the spatial components ($\alpha, \beta=x$, $y$) in the present
formalism, and the lattice gauge fields $A_{IJ}^{s}\equiv A_{\alpha }^{s}(I)$
($J=I-\hat{\alpha}$) and $A_{ij}^{h}\equiv A_{\alpha }^{h}(j)$ ($i=j+\hat{%
\alpha}$), with the indices $i$, $I$ standing for a square lattice site and
its dual lattice site, respectively \cite{kou2005}.

In this theory, the order parameter $\Delta _{ij}^{s}=\sum_{\sigma
}\left\langle b_{i\sigma }e^{-i\sigma A_{ij}^h}b_{j-\sigma
}\right\rangle \neq 0$ in $L_{s}$ [Eq.(\ref{Ls})] characterizes
the short-range bosonic RVB pairing onset at an upper pseudogap
temperature $T_{0}$. At $T<T_{0}$, the AF correlations start to
develop,\cite{gu2005} concomitant with the bosonic RVB pairing
condensation, as described by $L_{s}$. This theory
predicts\cite{kou2005} two ordered phases in the lower temperature
regions at $T\ll T_{0}$, namely, the AF long range order (AFLRO)
phase with spinon condensation and holon self-localization at low
doping, and the SC phase with holon condensation and spinon
confinement at higher doping.

In the backdrop of the RVB pairing and holon condensation, the SC
phase coherence is realized when spinons are confined in pairs. The
opposite case is also allowed, in which the spinons are not
confined, namely, single spinons are present as {\em free} neutral
objects. This defines the regime known as the spontaneous vortex
phsae (SVP) or lower pseudogap phase (LPP), previously discussed
based on the phase string model in the Hamiltonian formalism at a
generalized mean-field (Ginzburg-Landau)
level.\cite{weng2002,weng2006}

In the present work, we will reformulate the description of the SVP
based on the above mutual Chern-Simons gauge theory. Since the gauge
fluctuations beyond the mean-field level can be more faithfully
incorporated and treated in this Lagrangian approach, the present
formalism will be more suitable for describing the transitions from
the SVP to, e.g., the SC phase, as well as the nonlinear effects
like magnetic field dependence of magnetization, etc.


\subsection{Description of Spontaneous Vortex Phase}

The SVP is defined as the holon condensed phase, in which the
holong field can be decomposed as
$h_{I}=\sqrt{n_{I}^{h}}e^{i\theta _{I}}$. Noting that the
amplitude fluctuation $n_I^h$ of $h_I $ is gapped, with the phase
fluctuation $\theta_I$ as the most relevant mode, the Lagrangian
$L_{h}$ in Eq.(\ref{Lh}) can be approximately reexpressed, up to a
constant, as
\begin{widetext}
\begin{eqnarray}
L_{h} &=&\sum_{I}in_{I}^{h}\left[ \partial _{0}\theta
_{I}-A_{0}^{s}(I)-eA_{0}^{e}(I)\right] +\frac{u }{2}\sum_{I}\left(
n_{I}^{h}-\bar{n}^{h}\right) ^{2}
-2t_{h}\bar{n}^{h}\sum_{I(\hat{\alpha}=\hat{x},\hat{y})}\cos \left[
\theta _{I}-\theta _{I-\hat{\alpha}}-A_{\alpha
}^{s}(I)-eA_{\alpha }^{e}(I)\right] \nonumber \\
&\simeq &\sum_{I}in_{I}^{h}\left[ \partial _{0}\theta
_{I}-A_{0}^{s}(I)-eA_{0}^{e}(I)\right] +\frac{u }{2}\sum_{I}\left(
n_{I}^{h}-\bar{n}^{h}\right)
^{2}+t_{h}\bar{n}^{h}\sum_{I(\hat{\alpha}=\hat{x},\hat{y})}\left[
\Delta_{\alpha}\theta _{I}-A_{\alpha }^{s}(I)-eA_{\alpha
}^{e}(I)-2\pi N_{\alpha }(I)\right] ^{2} \nonumber
\end{eqnarray}
\end{widetext}in which $A_{\mu }^{e}$ is the external electromagnetic gauge
vector and $\bar{n}^{h}=\delta $. Note that in obtaining the last
line of the above expression, the following Villain approximation
has been used in the partition function
\begin{eqnarray}
& &e^{\gamma\cos \left[ \theta _{I}-\theta _{I-\hat{\alpha}}-A_{\alpha
}^{s}(I)-eA_{\alpha }^{e}(I)\right]} \simeq \mathrm{const.} \nonumber \\
& &\times\sum_{\left\{ N_{\alpha }\in \mathbb{Z}\right\} } e^{ -\frac{\gamma}{%
2}\left[ \theta _{I}-\theta _{I-\hat{\alpha}}-A_{\alpha }^{s}(I)-eA_{\alpha
}^{e}(I)-2\pi N_{\alpha }(I)\right] ^{2}}
\end{eqnarray}
at large $\gamma$ or low temperature. Consequently the resulting
effective holon Lagrangian becomes quadratic in $A_{\mu }^{s}$.

Since the mutual Chern-Simons term is linear in $A_{\mu }^{s}$ while the
spinon part is independent of $A_{\mu }^{s},$ one can thus integrate out $%
A^{s}$ in the total partition function to obtain the effective dual
Lagrangian $L_{\mathrm{dual}}^{h}$ as:
\begin{eqnarray}
L_{\mathrm{dual}}^{h} &=&\frac{u }{2\pi ^{2}}\sum_{i}\left[ B^{h}(i)-\pi
\bar{n}^{h}\right] ^{2}+\frac{1}{4\pi ^{2}t_{h}\bar{n}^{h}}\sum_{i,\alpha }{%
E_{\alpha }^{h}(i)}^{2}  \nonumber \\
&&-2i\sum_{i,\mu }A_{\mu }^{h}J_{\mu }^{\mathrm{vor}}-i\frac{e}{\pi }%
\sum_{i}\epsilon ^{\mu \nu \tau }A_{\mu }^{e}\Delta _{\nu }A_{\tau
}^{h} \label{Ldual}
\end{eqnarray}%
where $J_{0}^{\mathrm{vor}}\equiv \epsilon ^{\alpha \beta }\Delta _{\alpha
}N_{\beta }\in \mathbb{Z}$ is the temporal component of the current for the
vortices of the holon condensate and $J_{\alpha }^{\mathrm{vor}}\equiv
-\epsilon ^{\alpha \beta }\partial _{0}N_{\beta }$ denotes the spatial
component, which satisfy the conservation equation $\partial _{0}J_{0}^{%
\mathrm{vor}}+\Delta _{\alpha }J_{\alpha }^{\mathrm{vor}}=0$.

The above procedure leading to Eq. (\ref{Ldual}) is similar to the standard
boson-vortex duality transformation \cite{fisher1989}. Here the field
strengths of the Maxwell gauge field $A_{\mu }^{h}$ are represented by $%
B^{h}(i)=\epsilon ^{\alpha \beta }\Delta _{\alpha }A_{\beta }^{h}(i)$ and ${%
E_{\alpha }^{h}(i)=\partial }_{0}A_{\alpha }^{h}(i)-\Delta _{\alpha
}A_{0}^{h}(i)$ and the Gaussian fluctuations of $B^{h}(i)-\pi \bar{n}^{h}$
and ${E_{\alpha }^{h}(i)}$ are determined by the Maxwell terms in Eq. (\ref%
{Ldual}). Note that the integration over $A_{0}^{s}(I)$ results in
the same constraint as in Eq. (\ref{constraint1}): $B^{h}(i)=\pi
n_{I}^{h}$ which will be further constrained to $\pi \bar{n}^{h} $
in the limit of $u \rightarrow \infty $ when no density fluctuations
are allowed. Due to such a background ``magnetic'' field
$B^{h}\simeq \pi \bar{n}^{h}$, the spin state as governed by $L_{s}$
will be driven \cite{chen2005} into a spin liquid phase at finite
doping concentration $\bar{n}^{h}>0$ as opposed to an AFLRO ground
state at half-filling.

The effective Lagrangian $L_{\mathrm{dual}}^{h}$ also describes
the vortices of the holon condensate as ``charge'' $\pm 2$
particles interacting with the Maxwell gauge field $A_{\mu }^{h}$ via the minimal coupling $%
A_{\mu }^{h}J_{\mu }^{\mathrm{vor}}$. By contrast, the spinons in $L_{s}$
also act as the source of the gauge field $A_{\mu }^{h}$ with a charge $%
\sigma =\pm 1$. This corresponds to the fact that in the original
holon
language, each vortex with $J_{0}^{\mathrm{vor}}=\pm 1$ has a phase winding $%
\pm 2\pi $, while each spinon carries a half-vortex with a phase winding $%
\pm \pi ,$ known as the spinon-vortex. \cite{weng2002}
Energetically one expects that the charge-$2$ vortices cost more
than the charge-$1$ spinon-vortices and thus, to leading order
approximation, the former excitations can be neglected unless when
the temperature is very close to the upper boundary of the SVP
where the phase winding starts to lose rigidity.

But before dropping the term of $A_{\mu }^{h}J_{\mu }^{\mathrm{vor}}$ from $%
L_{\mathrm{dual}}^{h}$, one has to be careful about the case when a
$\pm 2\pi $ vortex is bound to a $\mp \pi $ spinon-vortex, resulting
a $\pm \pi $ vortex which is energetically the same as the original
spinon-vortex. Generally, including both $L_{s}$ and
$L_{\mathrm{dual}}^{h}$, the total
charge coupled to $A_{\alpha }^{h}(i)$ is given by $q_{i}^{\mathrm{tot}%
}=n_{i\uparrow }^{b}-n_{i\downarrow }^{b}+2J_{0}^{\mathrm{vor}}(i)$.
Consequently, the total charge can be $q_{i}^{\mathrm{tot}}=-1$ if there is
an up spinon co-existent with an anti-vortex of the holon condensate at site
$i$, \emph{i.e.,} $n_{i\uparrow }^{b}=1$ ($n_{i\downarrow }^{b}=0$) and $%
J_{0}^{\mathrm{vor}}(i)=-1$. Similarly a down spinon can carry a
$q^{\mathrm{tot}}=+1$
total charge by binding with an vortex with $J_{0}^{%
\mathrm{vor}}=1$. By defining $\Phi _{i}^{\dagger }$ and $\Phi _{i}$
as the rising and lowering operators of the holon vortices, with the
commutation relations $\left[ \Phi
_{i},J_{0}^{\mathrm{vor}}(j)\right] =\Phi _{i}\delta_{ij}$ and
$\Phi_i^\dagger\Phi_i=1$, $\left[ \Phi _{i},\Phi _{j}^{\dagger
}\right] =0$, the creation operators of the above-discussed
spinon-vortex
bound states can be written as $b_{i\uparrow }^{\dagger }\Phi _{i}$ and $%
b_{i\downarrow }^{\dagger }\Phi _{i}^{\dagger }$. Then, at each site, there
will be 4 states that carry the minimal total topological charge $q_{i}^{%
\mathrm{tot}}=\pm 1$:
\begin{eqnarray}
b_{i\uparrow }^{\dagger }\left\vert 0\right\rangle &=&\left\vert i\uparrow
+\right\rangle ,\qquad b_{i\downarrow }^{\dagger }\left\vert 0\right\rangle
=\left\vert i\downarrow -\right\rangle  \nonumber \\
b_{i\uparrow }^{\dagger }\Phi _{i}\left\vert 0\right\rangle &=&\left\vert
i\uparrow -\right\rangle ,\qquad b_{i\downarrow }^{\dagger }\Phi
_{i}^{\dagger }\left\vert 0\right\rangle =\left\vert i\downarrow
+\right\rangle  \label{fourtype}
\end{eqnarray}
where the sign $\pm $ denotes that of $q_{i}^{\mathrm{tot}}$. The
energy of each state above is composed of the superfluid vortex
energy and the spinon excitation. Since the superfluid energy only
depends on the total dual charge, $\left\vert
q^{\mathrm{tot}}\right\vert =1$, the four types of spinon-vortices
in Eq. (\ref{fourtype}) are degenerate in energy. In other words,
the spinon-vortex states defined in Eq. (\ref{fourtype}), rather
than the two-component spinons $b_{i\sigma }$ and the holon vortex
field $J_{\mu }^{\mathrm{vor}}$, describe the true low-energy
spin/vortex excitations in the SVP. Similar conclusion has been also
reached previously in the different mean-field (Ginzburg-Landau)
approach for the SVP.\cite{weng2006}

Based on the above discussion, by introducing $\left( \bar{b}_{i\uparrow },%
\bar{b}_{i\downarrow }\right) =\left( b_{i\uparrow }\Phi
_{i}^\dagger,b_{i\downarrow }\Phi _{i}\right) $ to stand for the two
new spinon-vortex states in Eq. (\ref{fourtype}) and neglect
independent $\pm 2\pi $ vortices, the low-lying effective theory of
the SVP in the mutual Chern-Simons gauge
theory description can be finally written down as%
\begin{equation}
L_{\mathrm{eff}}=\tilde{L}_{s}+\tilde{L}_{\mathrm{dual}}^{h}  \label{eff}
\end{equation}%
where
\begin{widetext}
\begin{eqnarray}
\tilde{L}_s&=&\sum_{i\sigma }b_{i\sigma }^{\dagger }\left[
\partial _{0}-i\sigma A_{0}^{h}(i)\right] b_{i\sigma }+\sum_{i\sigma }\bar{b}_{i\sigma }^{\dagger }\left[
\partial _{0}+i\sigma A_{0}^{h}(i)\right] \bar{b}_{i\sigma }-\frac J2\sum_{\left\langle ij\right\rangle
\sigma}\Delta^s_{ij}\left(b_{i\sigma}^\dagger b_{j-\sigma}^\dagger
e^{i\sigma A_{ij}^h}+\bar{b}_{i\sigma}^\dagger
\bar{b}_{j-\sigma}^\dagger e^{-i\sigma
A_{ij}^h}+h.c.\right)\nonumber\\
&+&\lambda\left[\sum_{i\sigma}\left(b_{i\sigma}^\dagger
b_{i\sigma}+\bar{b}_{i\sigma}^\dagger
\bar{b}_{i\sigma}\right)-N(1-\delta)\right]+\frac
12g\mu_BB^e\sum_{i\sigma}\sigma\left(b_{i\sigma}^\dagger
b_{i\sigma}+\bar{b}_{i\sigma}^\dagger \bar{b}_{i\sigma}\right) \label{Lsfinal}\\
\tilde{L}^h_{\rm dual}&=&\frac {u}{2\pi^2}\sum_i\left(
B^h(i)-\pi\bar{n}^h\right)^2+\frac1{4\pi^2
t_h\bar{n}^h}\sum_{i,\alpha}{E_{\alpha}^h}(i)^2-i\frac e\pi
\sum_i\epsilon^{\mu\nu\tau}A_\mu^e\Delta_\nu
A_\tau^h\label{Lhdualfinal}.
\end{eqnarray}
\end{widetext}in which the Zeeman coupling between the spinon and external
magnetic field $B^{e}$ is also included, with $g$ as the Lande g-factor and $%
\mu _{B}$ the Bohr magneton. Such an effective theory describes
4-flavor spinons minimally coupled to the Maxwell gauge field.
There will be two possible phases as the consequences: a
\emph{confined} phase in which all spinon-vortices form short-range
neutral pairs and a \emph{deconfined} phase in which they are free.
Physically, the former corresponds to the superconducting phase as
discussed in Ref. \cite{kou2005} and the latter is a spin liquid and
at the same time a vortex liquid, which corresponds to the SVP to be
further explored in this work.

\section{Mean-Field Approximation and physical properties}

\subsection{Mean-field Approximation}

The physical properties of the SVP is described by the effective
Lagrangian (\ref{eff}), in which $\tilde{L}_{s}$ in Eq. (\ref%
{Lsfinal}) determines the spinon degrees of freedom minimally coupled to the
gauge field $A_{\mu }^{h}$, while the latter is governed by $\tilde{L}_{%
\mathrm{dual}}^{h}$ in Eq. (\ref{Lhdualfinal}). Note that the external
electromagnetic field $A_{\mu }^{e}$ does not directly couple to the spinons
except for the Zeeman coupling, indicating that the spinons indeed do not
carry electromagnetic charge. On the other hand, $A_{\mu }^{e}$ and $A_{\mu
}^{h}$ are coupled by a mutual Chern-Simons term in Eq. (\ref{Lhdualfinal}),
implying that the spinons do carry superfluid vortices, which will become
clear in the following mean-field solution.

$\tilde{L}_{\mathrm{dual}}^{h}$ indicates the following mean-field solution
for $A_{\alpha }^{h}:$

\begin{equation}
\bar{B}^{h}(i)=\pi \bar{n}^{h}\text{, \ \ }{\bar{E}_{\alpha }^{h}(i)=0}\text{%
\ \ \ }
\end{equation}%
while $\bar{A}_{0}^{h}$ couples to the external magnetic field by $-iN\frac{e%
}{\pi }\bar{A}_{0}^{h}B^{e}$. To the leading order approximation,
the fluctuations of $A_{\alpha }^{h}$ and $A_{0}^{h}$ around
$\bar{A}_{\alpha }^{h}$ and $\bar{A}_{0}^{h}$ may be neglected here
if the temperature is not very close to $T_{c}$ or $T_{v}$, namely,
the phase boundaries. When $T$ approaches $T_{c}$, for example, the
long-range interaction between spinons mediated via the gauge field
$A_{\mu }^{h}$ will become important and the fluctuation of
${E_{\alpha }^{h}}$ in $\tilde{L}_{\mathrm{dual}}^{h}$ can no longer
be omitted \cite{kou2005}. We shall leave the discussion of the
phase boundaries to the next section. In the following, we first
focus on the physical consequences of the SVP at the mean-field
level, where the spinon-vortices proliferate and the long-range
interaction between them is well screened.

The effective Lagrangian (\ref{eff}) is then reduced to a mean-field
Lagrangian
\begin{equation}
L_{\mathrm{eff}}^{MF}=\tilde{L}_{s}(\Delta ^{s},\bar{A}_{\mu
}^{h},\lambda )-iN\frac{e}{\pi }\bar{A}_{0}^{h}B^{e}  \label{Lmf}
\end{equation}%
which can be diagonalized to $L_{\mathrm{eff}}^{MF}=\sum_{m\sigma
}\left[ \gamma _{m\sigma }^{\dagger }\left( \partial _{\tau
}+E_{m\sigma }\right) \gamma _{m\sigma }+\bar{\gamma}_{m\sigma
}^{\dagger }\left( \partial _{\tau }+\bar{E}_{m\sigma }\right)
\bar{\gamma}_{m\sigma }\right] +\mathrm{constant}$ by the standard
Bogoliubov transformation\cite{weng1998}
\begin{eqnarray}
b_{i\sigma } &=&\sum_{m}w_{m\sigma }(i)\left( u_{m}\gamma _{m\sigma
}-v_{m}\gamma _{m-\sigma }^{\dagger }\right)  \nonumber \\
\bar{b}_{i\sigma } &=&\sum_{m}\bar{w}_{m\sigma }(i)\left( u_{m}\bar{\gamma}%
_{m\sigma }-v_{m}\bar{\gamma}_{m-\sigma }^{\dagger }\right)
\end{eqnarray}%
with
\begin{eqnarray}
E_{m\sigma } &=&E_{m}+\sigma \left( \frac{g\mu _{B}B^{e}}{2}-i\bar{A}%
_{0}^{h}\right)  \nonumber \\
\bar{E}_{m\sigma } &=&E_{m}+\sigma \left( \frac{g\mu _{B}B^{e}}{2}+i\bar{A}%
_{0}^{h}\right)\nonumber\\
E_m&=&\sqrt{\lambda^2-\xi_m^2}
\end{eqnarray}%
and
\begin{eqnarray}
u_{m}=\sqrt{\frac{\lambda +E_{m}}{2E_{m}}},\text{  }v_{m}=\mathrm{sgn}(\xi _{m})\sqrt{%
\frac{\lambda -E_{m}}{2E_{m}}}
\end{eqnarray}
The wave function $\bar{w}_{m\sigma }(i)=w_{m\sigma }^{\ast }(i),$ and $%
w_{m\sigma }(i)$ satisfies the Schr$\ddot{\mathrm{o}}$dinger equation $-%
\frac{J}{2}\Delta ^{s}\sum_{j=\mathrm{nn(}i)}e^{i\sigma \bar{A}%
_{ij}^{h}}w_{m\sigma }(j)=\xi_{m}w_{m\sigma }(i)$.

By minimizing $L_{\mathrm{eff}}^{MF}$ with regard to the parameters
$\lambda $, $\Delta_{ij}^s=\Delta ^{s}$, and $\bar{A}_{0}^{h}$, the
following self-consistent equations are obtained
\begin{eqnarray}  \label{vortexnu}
1-\delta&=&\frac{1}{N}\sum_{i,\sigma }\left\langle b_{i\sigma }^{\dagger }b_{i\sigma }+%
\bar{b}_{i\sigma }^{\dagger }\bar{b}_{i\sigma }\right\rangle
\label{spinonnu} \\
\Delta ^{s}&=&\sum_{\sigma }\left\langle  b_{j-\sigma
}e^{i\sigma\bar{A}_{ji}^{h}
}b_{i\sigma }+\bar{b}_{j-\sigma }e^{-i\sigma\bar{A}_{ji}^{h} }\bar{b}%
_{i\sigma } \right\rangle \\
-\frac{B^{e}a^{2}}{\phi_{0}}&=&\frac{1}{N}\sum_{i\sigma }\sigma
\left\langle b_{i\sigma }^{\dagger }b_{i\sigma }-\bar{b}_{i\sigma
}^{\dagger }\bar{b}_{i\sigma }\right\rangle \label{vn}
\end{eqnarray}%
the first two of which self-consistently determine the average
spinon number and the RVB order parameter, and the last equation
describes the polarization of the total vorticity of the
spinon-vortices by the external magnetic field. Here and below,
$\phi_0=\frac{hc}{2e}$ stands for the flux quanta. Once $\lambda $,
$\Delta ^{s}$ and $\bar{A}_{0}^{h}$ are decided self-consistently,
physical properties of the SVP can be calculated straightforwardly.

\subsection{Magnetization}

The total magnetization can be decided by
\begin{equation}
M_{\mathrm{tot}}=-\frac{1}{V}\frac{\partial F}{\partial B^{e}}
\end{equation}%
with $F$ the mean-field free energy $F=-k_{B}T\ln \left[ \int d[b]d[b^{\ast
}]\exp (-\int_{0}^{\beta }L_{\mathrm{eff}}^{MF}d\tau )\right] $ and $%
V=Na^{2}d$ the average volume per $\mathrm{CuO_{2}}$ layer ($d$ is the
interlayer spacing). From Eqs. (\ref{Lmf}) and (\ref{Lsfinal}), the total
magnetization can be then expressed explicitly as
\begin{equation}
M_{\mathrm{tot}}=\frac{e}{\pi a^{2}d}i\bar{A}_{0}^{h}-\frac{g\mu _{B}}{a^{2}d%
}\left\langle S^{z}\right\rangle \equiv M_{\mathrm{dia}}+M_{\mathrm{para}}
\end{equation}%
in which $M_{\mathrm{dia}}\propto i\bar{A}_{0}^{h}$, $M_{\mathrm{para}%
}\propto -\left\langle S^{z}\right\rangle $ stand for the orbital
diamagnetism from the vorticies and the paramagnetism from the Zeeman
coupling, respectively, where $i\bar{A}_{0}^{h}$ is decided by the
self-consistent equations (\ref{vortexnu}). Compared to the conventional
vortex liquid theory, the origin of both the diamagnetism and paramagnetism
here is intrinsically related to the same spin degrees of freedom, which is
a unique feature of the present mutual Chern-Simons theory.

\begin{figure*}[tbp]
\begin{center}
\includegraphics[width=2.5in] {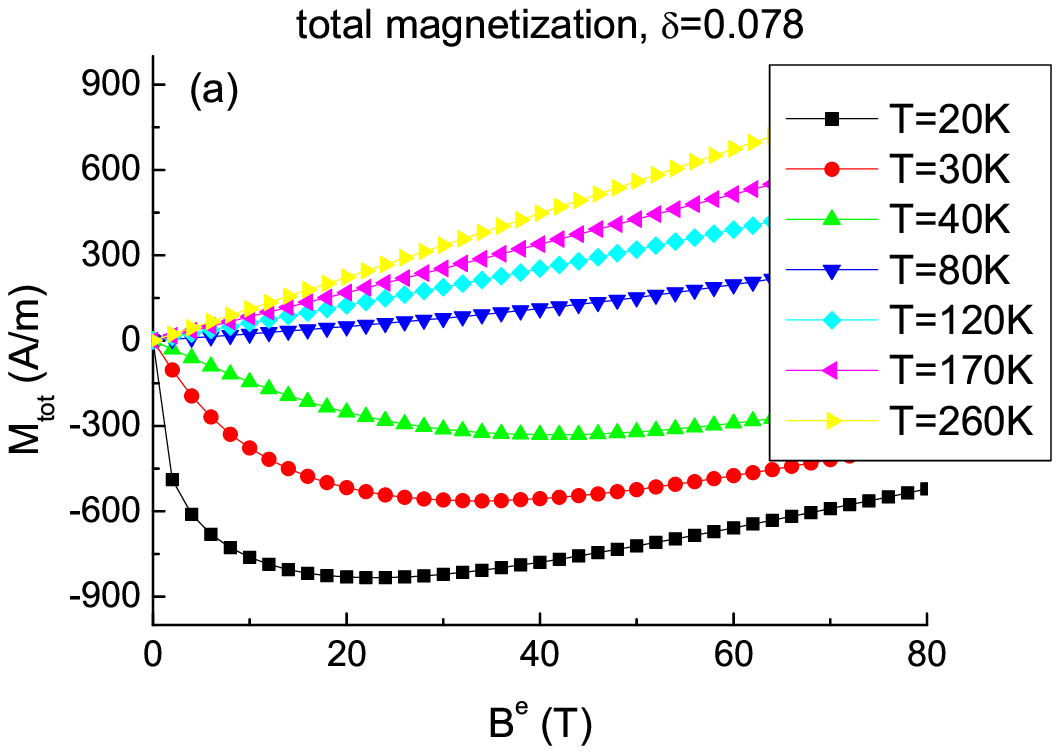}%
\includegraphics[width=2.5in]
{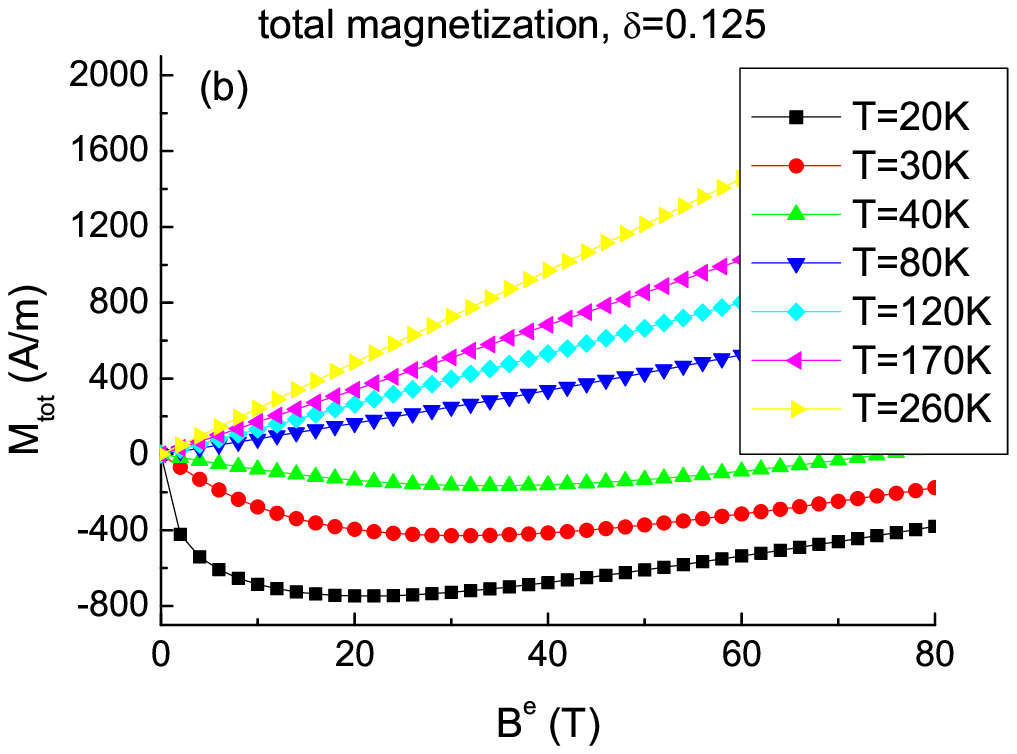}
\par
\includegraphics[width=2.5in] {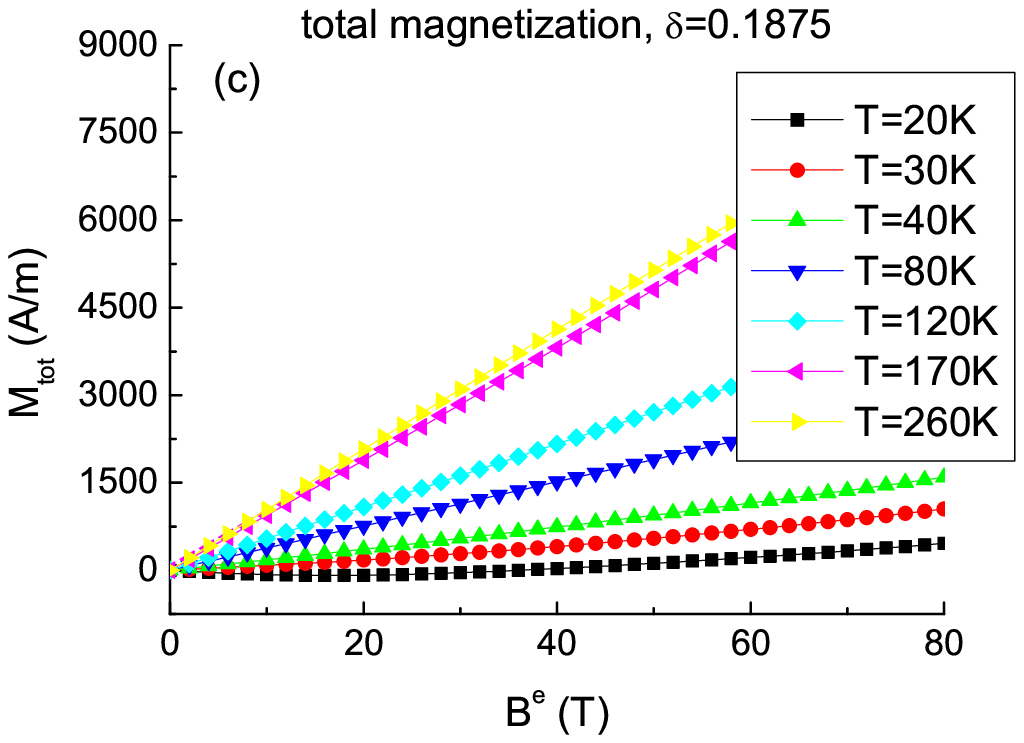}%
\includegraphics[width=2.5in]
{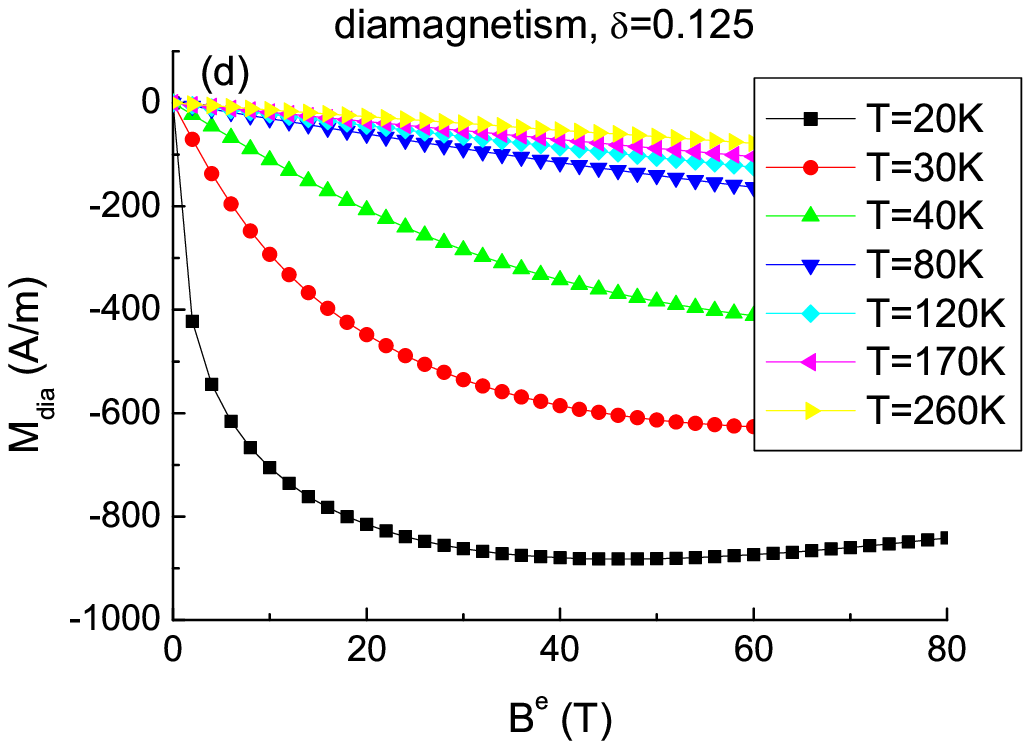}
\end{center}
\caption{The $B^e$ dependence of the total magnetization $M=M_{\mathrm{dia}%
}+M_{\mathrm{para}}$ at various doping concentrations: (a) $\protect\delta %
=0.078$, (b) $\protect\delta =0.125$, (c) $\protect\delta =0.188;$ (d) the
diamagnetism $M_{\mathrm{dia}}$ at $\protect\delta =0.125$. The parameters
in the mean-field theory are chosen as $a=5.5\mathring{A}$, $d=7.7\mathring{A%
}$, $J=120\mathrm{meV}$. } \label{magnetization}
\end{figure*}

\begin{figure}[tbp]
\begin{center}
\includegraphics[width=3in] {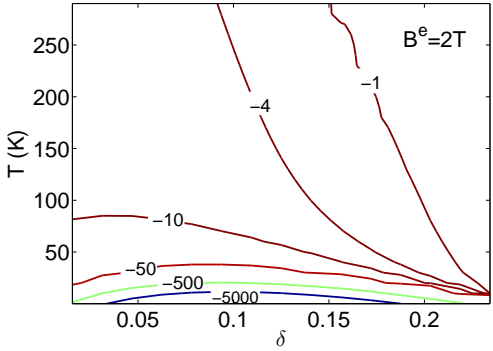}
\end{center}
\caption{The contour plot of the diamagnetism $M_{\rm dia}$ (in
units of ${\rm A/m}$) under a weak field $B^e=2T$.}
\label{contourMd}
\end{figure}

The magnetic field and temperature dependence of the total magnetization at
different doping concentrations as well as the diamagnetism part $M_{\mathrm{%
dia}}$ at $\delta =0.125$ are shown in Fig. \ref{magnetization} based on the
above mean-field theory. The contour plot of $M_{\mathrm{dia}}$ in the
temperature and doping space at $B^{e}=2$ $\mathrm{Tesla}$ is presented in
Fig. \ref{contourMd}. Note that the diamagnetism disappears at $\delta
\rightarrow 0$ where the density of the condensate vanishes, as well as at $%
\delta \rightarrow x_{\mathrm{RVB}}\simeq 0.25$ (see
Ref.\onlinecite{gu2005}) where the RVB pairing vanishes with the
proliferation of the unpaired spinons at $T=0$. The magnitude of
diamagnetism we obtained is comparable to the experimental
observations\cite{wang2005} in the weak-field region, but
over-estimated under strong field $B^e\sim H_{c2}$, where the upper
phase boundary of the SVP is reached and the mean-field
approximation is not applicable.

\begin{figure}[tbp]
\begin{center}
\includegraphics[width=3in] {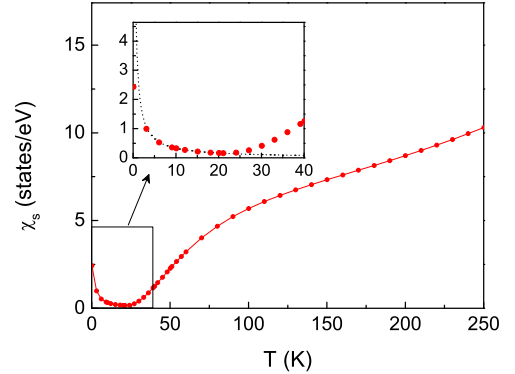}
\end{center}
\caption{The spin susceptibility
$\protect\chi_s=M_{\mathrm{para}}/B^e$ for doping
$\protect\delta=0.125$ and $B^e=2T$. Inset is the detail of low
temperature behavior. The dotted line stands for the Curie curve $\protect\chi_{%
\mathrm{c}}=\frac{n_s}{k_BT}$, with $n_s=B^ea^2/\phi_0$.}
\label{chis}
\end{figure}

The paramagnetic magnetization determines the spin paramagnetic
susceptibility
$\chi _{s}=M_{p}/B^{e}$ whose temperature dependence at $\delta =0.125$ and $%
B^{e}=2$ $\mathrm{Tesla}$ is shown in Fig. \ref{chis}. $\chi _{s}$
exhibits a prominent \textquotedblleft spin gap\textquotedblright\
behavior with reducing temperature below $100$ \textrm{K} similiar
to the pseudogap behavior in the zero field limit \cite{gu2005}.
Furthermore a Curie-type upturn emerges at lower temperature within
the \textquotedblleft spin gap\textquotedblright , which counts for
the contribution from the free moments in the vortex cores.
According to Eq. (\ref{vortexnu}), the total vorticity density
satisfies $n_{+}-n_{-}=B^{e}a^2/\phi_{0}$, and at low temperature
when the thermal excitations of spinons are suppressed, one finds
$n_{-}\simeq 0,n_{+}\simeq B^{e}a^2/\phi_{0}$, which means there are
on average $n_{s}=B^{e}a^2/\phi_{0}$ unpaired spinons per site. At
weak magnetic field $B^{e}\ll \phi_{0}/a^{2}$, $n_{s}\ll 1$ and the
unpaired spinons are very dilute and thus are nearly-independent to
each other. Consequently, a Curie-type spin susceptibility $\chi
_{\mathrm{c}}=n_{s}/k_{B}T$ is expected as shown in Fig. \ref{chis}.
Experimentally, the NMR spin-lattice relaxation rate for the
in-plane oxygen nuclear spins, which is related to $\chi _{s}$ due
to the hyperfine coupling constant, does show a Curie-type
temperature
behavior in the vortex cores in the superconducting phase\cite{mitrovic2003,kakuyanagi2003}%
, before it eventually decreases and vanishes at very low
temperature which is presumably due to the Kondo screening effect of
the free moments by the nodal quasiparticles.

\subsection{Magneto-Resistivity}

The charge resistivity is non-vanishing in the SVP, although the holon
condensation is still present. The origin of the dissipation is due to the
flow of the spinon-vortices, whose response to an external electric field
can be seen by taking the classical equation of motion for $A_{\alpha }^{h}$
in the dual Lagrangians, (\ref{Lsfinal}) and (\ref{Lhdualfinal}), $\frac{%
\partial L_{\mathrm{eff}}}{\partial A_{\alpha }^{h}}=0$, which leads to the
spinon-vortex current:
\begin{equation}
J_{\alpha }^{sv}\equiv i\frac{\partial \tilde{L}_{s}}{\partial
A_{\alpha
}^{h}}=\frac{e}{\pi }\epsilon _{\alpha \beta }E_{\beta }^{e}+i\left( \frac{1%
}{2\pi ^{2}t_{h}\bar{n}^{h}}\partial _{\tau }E_{\alpha
}^{h}-\frac{u}{\pi ^{2}}\epsilon _{\alpha \beta }\Delta _{\beta
}B^{h}\right)
\end{equation}%
In a uniform stationary state, the last two terms vanish and thus one
obtains
\begin{equation}
J_{\alpha }^{sv}=\frac{\epsilon _{\alpha \beta }cE_{\beta
}^{e}}{\phi _{0}} \label{JvEe}
\end{equation}%
after recovering the full units.

In a similar way, the equation of motion for $A_{i\alpha }^{s}$ based on the
\emph{original} Lagrangian (\ref{LMCS}) leads to the following relation
\begin{equation}
J_{\alpha }^{h}=-\frac{\epsilon _{\alpha \beta }cE_{\beta
}^{h}}{\phi _{0}} \label{JhEh}
\end{equation}%
where $J_{\alpha }^{h}\equiv ie\frac{\partial L_{h}}{\partial
A_{\alpha }^{s}} $ is the holon charge current which decides the
strength of
"electric" field $E_{\alpha }^{h}$ for the gauge field $A_{\alpha }^{h}$\cite%
{kou2005}.

Since the spinon-vortices directly see $A^{h}$ in the Lagrangian (\ref%
{Lsfinal}), there is generally a linear response relation between $\mathbf{E}%
^{h}$ and $\mathbf{J}^{sv}$:
\begin{eqnarray}
J_{\alpha }^{sv}=\sigma _{\alpha \beta }^{sv}E_{\beta
}^{h}\label{JvEh}
\end{eqnarray}
where $\sigma _{\alpha \beta }^{sv}$ denotes the spinon-vortex
conductivity. By combining (\ref{JvEe})-(\ref{JvEh}) and noting that
the holon current is equal to the electric current in the mutual
Chern-Simons theory, one finally obtains the electric resistivity as follows%
\[
\rho _{\alpha \beta }=-\left( \phi _{0}/c\right) ^{2}\epsilon
_{\alpha \gamma }\sigma _{\gamma \delta }^{sv}\epsilon _{\delta
\beta }
\]%
If one uses a simple semi-classical approximation by expressing the
spinon-vortex conductivity as $\sigma _{\alpha \beta }^{sv}\simeq \frac{%
n_{v}}{\eta _{s}}\delta _{\alpha \beta }$, with $n_{v}$ and $\eta _{s}$ the
total vortex number and viscosity, respectively, then the charge resistivity
in the SVP is given by
\begin{equation}
\rho \equiv \rho _{xx}=\frac{n_{v}}{\eta _{s}}\left( \frac{\phi _{0}}{c}%
\right) ^{2}  \label{rho}
\end{equation}

The resistivity in (\ref{rho}) is similar to the flux-flow resistivity in a
Type II superconductor except that $n_{v}$ in general is not simply
proportional to the external magnetic field $B^{e}$. Namely, in the SVP the
spinon-vortices can be spontaneously (thermally) generated with $n_{v}\neq 0,
$ such that $\rho \neq 0$ even at $B^{e}=0$. The resistivity $\rho (B^{e})$
can be expanded as
\begin{equation}
\rho (B^{e})=\rho (0)\left[ 1+\gamma B^{e2}+o(B^{e2})\right]
\end{equation}%
where the odd power terms of $B^{e}$ vanish due to the symmetry
$\rho (B^{e})=\rho (-B^{e})$. Suppose that the dependence of the
viscosity $\eta _{s}$ on $B^{e}$ is negligible, then the quadratic
coefficient $\gamma $ can be expressed as
\begin{equation}
\gamma =\frac{\rho (B^{e})-\rho (0)}{\rho (0)B^{e2}}\simeq \frac{%
n_{v}(B^{e})-n_{v}(0)}{n_{v}(0)B^{e2}}.  \label{nu}
\end{equation}

\begin{figure}[tbp]
\begin{center}
\includegraphics[width=3in] {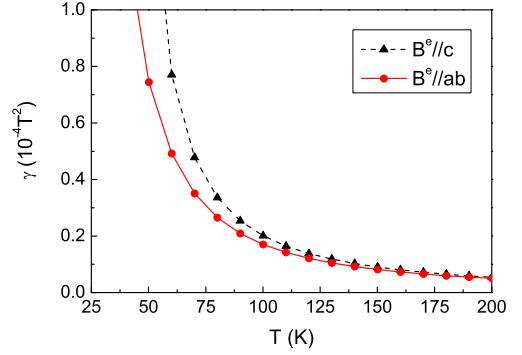}
\end{center}
\caption{The magento-resistence coefficient $\protect\gamma $ vs temperature
for the magnetic field $B^{e}$ which is either perpendicular or transverse
to the $\mathrm{ab}$ plane. }
\label{drho}
\end{figure}

The spinon-vortex density $n_{v}$ is determined in the mean-field
approximation by
\begin{eqnarray}
n_{v} &=&\frac{1}{N}\sum_{m,\sigma }\left\langle \gamma _{m\sigma }^{\dagger
}\gamma _{m\sigma }+\bar{\gamma}_{m\sigma }^{\dagger }\bar{\gamma}_{m\sigma
}\right\rangle   \nonumber \\
&=&\frac{1}{N}\sum_{m,\sigma }\left( \frac{1}{e^{\beta E_{m\sigma }}-1}+%
\frac{1}{e^{\beta \bar{E}_{m\sigma }}-1}\right)   \label{nv}
\end{eqnarray}%
With Eqs. (\ref{nu}) and (\ref{nv}), the coefficient $\gamma _{\perp }$ and $%
\gamma _{\parallel }$, with the external magnetic field $B^{e}$
perpendicular and parallel to the 2D plane, respectively, can be calculated
numerically as shown in Fig. \ref{drho}. An important prediction of the
present theory, as shown by Fig. \ref{drho}, is that $\gamma _{\Vert }$ is
\emph{comparable} to $\gamma _{\perp }$ in the SVP. This is a rather unusual
case for a vortex-flow-induced resistivity, since normally the in-plane
vortices are always created by the perpendicular magnetic field in a Type II
superconductor, where the vortex-flow-induced resistivity only exhibits
field-dependent magneto-resistivity for the compoenet of $B^{e}$ which is
perpendicular to the plane. Experimentally, the c-axis resistivity shows an
insulating behavior in the pseudo-gap phase until $T\sim T_{c}$ at low
doping, which implies that the interlayer quantum phase coherence is not
important. Thus a magnetic field parallel to the ab plane is not expected to
contribute significantly to the in-plane resistivity based on a conventional
flux-flow picture, which would predict $0\simeq \gamma _{\Vert }\ll \gamma
_{\perp }$.

But in the present theory, vortices are tied to the free spinons in the SVP.
Since the latter can be created by the Zeeman term with the external
magnetic field pointing at \emph{any} direction, the former can thus be
created by the in-plane field as well, although the total vorticity of the
2D orbital supercurrents still satifies the constraint (\ref{vn}) in which $%
B^{e}$ should be replaced by $B_{\perp }^{e}$. Although the
present mean-field result of $\rho $ may not be expected to be
quantitatively accurate in view of possible corrections from the
fluctuations, the existence of an \emph{anomalous} transverse
magneto-resistivity with $\gamma _{\Vert }$ comparable to $\gamma
_{\perp }$ remains a peculiar prediction based on the mutual
Chern-Simons theory, which is qualitatively consistent with the
experimental results for the underdoped
YBCO\cite{harris1995,ando2002}.

\subsection{Nernst Effect and Vortex Entropy}

The Nernst effect refers to a transverse electric field $E_{y}$
induced by applying a temperature gradient $-\nabla _{x}T$. The
Nernst coefficient is defined as
\begin{equation}
e_{y}=\frac{E_{y}}{-\nabla _{x}T}.
\end{equation}%
One important mechanism that can lead to a significant Nernst signal is the
flux flow in the vortex liquid phase, the contribution of which can be
calculated by using Eq. (\ref{JvEe}). Here the vorticity current driven by a
temperature gradient $-\nabla _{\alpha }T$ is
\begin{eqnarray}
J_{\alpha }^{sv}=\nu _{\alpha \beta }^{sv}\left( -\nabla _{\beta
}T\right)\label{vortexthc}
\end{eqnarray}
which, combined with Eq. (\ref{JvEe}), decides the Nernst coefficient
\begin{eqnarray}
e_{y}=\frac{\phi_0}{c}\frac{J_{x}^{v}}{-\nabla _{x}T}=\frac{\phi
_{0}}c\nu _{xx}^{sv}\nonumber
\end{eqnarray}

The vortex thermo-conductivity formula (\ref{vortexthc}) provides a
systematic way to calculate the Nernst coefficient. As in the last
section, we can introduce a simple drift approximation usually used
in a vortex liquid phase to obtain a leading order estimation of the
Nernst effect in the SVP, and leave the more microscopic calculation
to future works.

If the vortices drift with a velocity $v_{x}$ under the temperature
gradient, the vortex current $J_{x}^{sv}$ can be approximated by $%
J_{x}^{s-v}\simeq (n_{+}-n_{-})v_{x}=\frac{B^{e}}{\phi _{0}}v_{x}$,
in which the last equality comes from vorticity constraint
(\ref{vn}), and the magnetic field is understood as applied in the
perpendicular direction in the following. The velocity $v_{x}$ can
be decided by the equation $-s_{\phi }\nabla _{x}T=\eta _{s}v_{x}$,
in which $s_{\phi }$ is the transport entropy per vortex, and $\eta
_{s}$ is the same vortex viscosity as in the resistivity formula
(\ref{rho}). Thus we have $\nu_{xx}^{sv}=B^es_\phi/\phi_0\eta_s$,
which leads to the Nernst signal
\begin{equation}
e_{y}=\frac{B^{e}s_{\phi }}{c\eta _{s}} \label{ey}
\end{equation}

\begin{figure}[tbp]
\begin{center}
\includegraphics[width=3in] {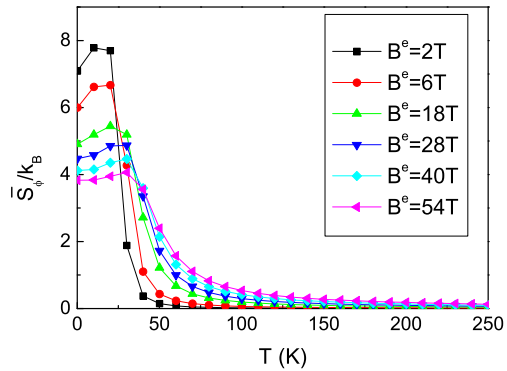}
\end{center}
\caption{The ratio $\bar{S}_\phi /k_{B}={\phi _{0}e_{y}}/%
\protect\rho {k_{B}}$ vs. temperature at $\protect\delta =0.125$, and
various magnetic field strengths. }
\label{entropy}
\end{figure}

According to Eqs. (\ref{rho}) and (\ref{ey}), the viscosity $\eta _{s}$ can
be eliminated by the ratio
\begin{equation}
\bar{S}_{\phi }\equiv \frac{\phi _{0}e_{y}}{c\rho }=\frac{B^{e}s_{\phi }}{%
\phi _{0}n_{v}}  \label{S}
\end{equation}%
which has the dimension of entropy and relates the transport entropy
and the density of vortices to the observables $e_{y}$ and $\rho $.
For the conventional Abrikosov vortex liquid, one has $B^{e}=\phi
_{0}n_{v}$ and thus $\bar{S}_{\phi }=s_{\phi }$, as usually used in
the analysis of
experiments\cite{capan2002}. However, in the present SVP, both $n_{+}$ and $%
n_{-}$ are non-vanishing such that generally $\bar{S}_{\phi }<s_{\phi }$. In
the mean-field theory, $s_{\phi }$ can be estimated by $s_{\phi }=S/n_{v}$,
in which $S$ is the entropy density of the spinons, and the numerical
results are presented in Fig. \ref{entropy} at different magnetic fields.
Note that such an estimation does not distinguish the transport entropy and
those which do not contribute to the transport (such as the vortex
configuration entropy), and may generally lead to an overestimate of $%
s_{\phi }$ and thus $\bar{S}_{\phi }$. Furthermore, $\bar{S}_{\phi
}$ should be further reduced due to the vortex-pinning effect at low
temperature in the superconducting phase, which is not included in
the present mean-field theory. Experimentally, the Nernst effect in
the SVP has been studied
systematically.\cite{xu2000,wang2002,capan2002,wang2003,wang2006} In
Ref. \onlinecite{capan2002}, the Nernst signal and
resistivity are measured for the underdoped $\mathrm{%
La_{1.92}Sr_{0.08}CuO_{4}}$, which shows $e_{y}\simeq 7\mathrm{\mu V/K}$ and
$\rho \simeq 2\times 10^{-6}\mathrm{\Omega \cdot m}$ at $H=10\mathrm{T}$, $T=25%
\mathrm{K}$. Taking the c-axis lattice constant $d\simeq
13\mathring{A}$ for LSCO\cite{harshman1992}, we obtain
$\bar{S}_{\phi }\simeq 0.7k_{B}$, which is about one order of
magnitude smaller than our estimation. Nevertheless, Fig.
\ref{entropy} provides an upper bound of the spinon transport
entropy, which drives the spinons downward the temperature gradient.

\section{Phase Boundaries of the Spontaneous Vortex Phase}

In the last section, the mean-field approximation of the mutual Chern-Simons
gauge theory is applied to describe the \emph{bulk} properties of the
spontaneous vortex phase. The fluctuations beyond the mean-field solution
will become important when the phase boundary of the SVP is considered.

The lower phase boundary of the SVP at low temperature corresponds
to the superconducting phase transition. With reducing temperature,
the number of thermally excited vortices in the SVP decreases and
the screening of the vortex interaction becomes weakened. Eventually
at $T\rightarrow T_{c}^{+}$ a vortex-antivortex confining transition
will take place and the system becomes SC phase coherent, which will
be discussed firstly below based on the mutual Chern-Simons theory.

\subsection{Superconducting transition}

In order to study the superconducting phase transition, let us start with
the general effective Lagrangian (\ref{eff}) of the SVP. For simplicity, we
shall consider the limit $u \rightarrow \infty $ which means that the holon
density fluctuations are not important and can be neglected. In this limit
the phase transition will be solely driven by vortices as expected. Under
such an approximation, the spatial component $A_{\alpha }^{h}$ cannot
fluctuate and is constrained by%
\begin{equation}
B^{h}=\pi \bar{n}^{h}.
\end{equation}%
Thus the only important dynamical variable will be the temporal component $%
A_{0}^{h}$.

By integrating out $b_{i\sigma },\bar{b}_{i\sigma }$ in Lagrangian (\ref{eff}%
)
\begin{equation}
e^{-S_{\mathrm{eff}}[A_{0}^{h}]}\equiv \int D[b_{i\sigma }]D[\bar{b}%
_{i\sigma }]\cdot \exp \left[ -\int_{0}^{\beta }d\tau L_{\mathrm{eff}}^{{}}%
\right]
\end{equation}%
one obtains a classical action for $A_{0}^{h}$ in uniform external field $%
\mathbf{\hat{z}\cdot }(\nabla \times \mathbf{A}^{e})=B^{e}$ and $\mathbf{E}%
^{e}=0$
\begin{eqnarray}
S_{\mathrm{eff}}[A_{0}^{h},B^{e}] &=&\beta \left( F_{s}[A_{0}^{h},B^{e}]-i%
\frac{e}{\pi }\sum_{i}A_{0}^{h}(i)B^{e}\right.  \nonumber \\
&&\left. +\frac{1}{4\pi ^{2}t_{h}\bar{n}^{h}}\sum_{i\alpha }\left( \Delta
_{\alpha }A_{0}^{h}(i)\right) ^{2}\right)  \label{freeenergy}
\end{eqnarray}%
in which $e^{-\beta F_{s}}\equiv \int D[b_{i\sigma }]D[\bar{b}_{i\sigma
}]\exp \left( -\int_{0}^{\beta }d\tau \tilde{L}_{s}\right) $.

Under the standard RPA approximation, $F_{s}$ can be expanded to the
quadratic order of $A_{0}^{h}$ based on Eq. (\ref{Lsfinal}):
\begin{equation}
F_{s}[A_{0}^{h},B^{e}]\simeq -iA_{0}^{h}(\mathbf{q=0)}N_{v}+\frac{1}{2}\sum_{%
\mathbf{q}}\chi _{\mathbf{q}}(B^{e})A_{0}^{h}(\mathbf{q})A_{0}^{h}(-\mathbf{q%
})  \label{RPAfreeenergy0}
\end{equation}%
in which the total vortex number $N_{v}=\sum_{i}\left\langle
n_{v}(i)\right\rangle $ with $n_{v}(i)\equiv \sum_{\sigma }\sigma \left(
b_{i\sigma }^{\dagger }b_{i\sigma }-\bar{b}_{i\sigma }^{\dagger }\bar{b}%
_{i\sigma }\right) $ and the susceptibility $\chi _{\mathbf{q}}$ is defined
by
\begin{equation}
\chi _{\mathbf{q}}=\frac{1}{N}\int_{0}^{\beta }\sum_{i,j}e^{i\mathbf{q\cdot
(r_{i}-r_{j})}}\left( \left\langle n_{v}(i,\tau )n_{v}(j,0)\right\rangle
-\left\langle n_{v}\right\rangle \left\langle n_{v}\right\rangle \right)
\label{chiq}
\end{equation}%
which is equal to the static spin susceptibility according to the
spinon-vortex binding. Note that here the dynamic part of
$A_{0}^{h}(i\omega _{n})$ for $\omega _{n}\neq 0$ is omitted. By
further using the expansion
\begin{equation}
\chi _{\mathbf{q}}\simeq \chi _{0}\left( 1+\lambda_0 q^{2}\right)
\end{equation}%
in the long-wavelength limit, one finds
\begin{widetext}
\begin{equation}
F_{s}[A_{0}^{h},B^{e}]\simeq -iA_{0}^{h}(\mathbf{q=0)}N_{v}+\frac{\chi _{0}}{%
2}\left( \sum_{i}\left( {A_{0}^{h}(i)}\right) ^{2}+\lambda_0
\sum_{i\alpha }\left( \Delta _{\alpha }A_{0}^{h}\right) ^{2}\right)
. \label{RPAfreeenergy}
\end{equation}
\end{widetext}

However, by noting that the totoal vortex number $N_{v}$ is an integer and
is conserved in the system described by Lagrangian (\ref{Lsfinal}), the
exact free energy $F_{s}$ should be a periodical function of $%
A_{0}^{h}(i) $ in the following sense
\begin{equation}
\exp \left\{ -\beta F_{s}[A_{0}^{h}+\frac{2\pi }{\beta },B^{e}]\right\}
=\exp \left\{ -\beta F_{s}[A_{0}^{h},B^{e}]\right\}  \label{periodic}
\end{equation}%
Consequently, the correct form of $F_{s}$ satisfying (\ref{periodic}) as
well as (\ref{RPAfreeenergy}) in the weak limit of ${A_{0}^{h}(i)}$ should
have the compact version
\begin{eqnarray}
F_{s}[A_{0}^{h},B^{e}]&\simeq& i\frac{e}{\pi }\sum_{i}A_{0}^{h}B^{e}+\frac{1%
}{2}\chi _{0}\lambda_0 \sum_{i\alpha }\left( \Delta _{\alpha
}A_{0}^{h}\right)
^{2}  \nonumber \\
& &-\frac{1}{\beta ^{2}}\chi _{0}\sum_{i}\cos \left( \beta A_{0}^{h}\right)
\label{Compactfreeenergy0}
\end{eqnarray}%
in which the self-consistent equation (\ref{vortexnu}) for $N_{v}$
is used, and $\chi _{0},\lambda_0 $ can be calculated based on Eq.
(\ref{Lsfinal}).

Finally, combining Eqs. (\ref{freeenergy}) and (\ref{Compactfreeenergy0}),
and redefining $\beta A_{0}^{h}(i)=\phi (\mathbf{r}_{i})$, we find that $S_{%
\mathrm{eff}}$ reduces to a sine-Gordon action
\begin{equation}
S_{\mathrm{eff}}\simeq \int d^{2}\mathbf{r}\left\{ \frac{K}{2}\left( \nabla
\phi \right) ^{2}-2y\cos \phi \right\}  \label{sinegordon}
\end{equation}%
with
\begin{eqnarray}
K &\equiv &\beta ^{-1}\left( \frac{1}{2\pi ^{2}t_{h}\bar{n}^{h}}+\chi
_{0}\lambda_0 \right) ,  \label{K} \\
y &\equiv &\frac{\chi _{0}(B^{e})}{2\beta }.  \label{y}
\end{eqnarray}%
Such a sine-Gordon effective action describes the fluctuations beyond the
mean-field theory at low temperature where the holon density fluctuations
are negligible and holon $2\pi $ vortices are not important.

In the following, we discuss the superconducting phase transition
based on this action at low temperature, where the phase boundary
of the SVP is characterized by the superconducting temperature
$T_{c}$ and melting magnetic field $H_{m}$.

First, consider the case at $B^{e}=0$. If $\chi _{0}$ is negligible (the
suppression of spin fluctuations) and $y\simeq 0$, the Kosterlitz-Thouless
(KT) phase transition temperature is decided by the universal value
\begin{eqnarray}
K=\frac{1}{8\pi }\Rightarrow T_{\mathrm{KT}}=\frac{\pi }{4}t_{h}\bar{n}^{h}
\label{TKT}
\end{eqnarray}%
which reproduces the conventional KT physics of the holon condensate
when the long-wavelength spin fluctuations are absent. Note that
$T_{\mathrm{KT}}$ in this limit is only $1/4$ of the usual XY model
with the same stiffness $t_{h}\bar{n}^{h}$ because the spinon
vortices are $\pi $-vortices instead of conventional $2\pi $
vortices and the latter are neglected at low temperature.

\begin{figure}[tbp]
\begin{center}
\includegraphics[width=3.2in] {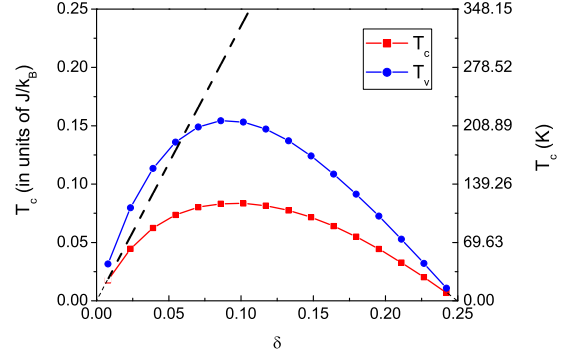}
\end{center}
\caption{Doping dependence of $T_c$ decided by Eq.
(\ref{selfconsistentTc}) and $T_v$ decided by Eq. (\ref{nv=delta}),
in which the dashed line stands for bare KT transition temperature
$T_{KT}$ in Eq. (\ref{TKT}). All the calculations are done for
$t_h=3J$. } \label{plotTvHc2}
\end{figure}

However, the spin susceptibility $\chi _{0},$ although suppressed at
low temperature in the present pseudogap phase, can be greatly
enhanced at the temperature comparable to the spin gap energy
(\textit{cf.} Fig. \ref{chis}), which
can then suppress the phase transition temperature from the universal value $%
T_{\mathrm{KT}}$ if the latter is higher than the characteristic spin gap
scale. To the linear order of $y$ and $K$, the equation of critical line on
the $K-y$ plane can be written as\cite{kogut1979}
\begin{equation}
y=2(1-8\pi K)
\end{equation}%
which results in
\begin{equation}
T_{c}=\frac{1}{{\chi _{0}(T_{c})}\left( \frac{1}{4}+8\pi \lambda_0 \right) +%
\frac{4}{\pi t_{h}\bar{n}^{h}}}  \label{selfconsistentTc}
\end{equation}%
This is a self-consistent equation for $T_{c}$ as ${\chi _{0}(T)}$ is
strongly temperature dependent as shown in Fig. \ref{chis}. The numerical
solution of the superconducting transition temperature is presented in Fig. %
\ref{plotTvHc2}, whose value is comparable with the experimental
results. By contrast, the corresponding unrenormalized
$T_{\mathrm{KT}}$ is shown in the same figure for comparison. It
is should be noted that in the cuprate superconductors, the
superconducting transition can deviate from a pure 2D KT
transition \cite{kamal1994,osborn2003} due to quantum
fluctuations ($\emph{e.g.,}$\emph{\ }the dynamic fluctuations of $A_{0}^{h}$%
) as well as the interlayer coupling which may lead to a 3D critical
behavior near $T_{c}$. Nevertheless, the present mutual Chern-Simons theory
clearly illustrates that the low-lying spin correlations play a crucial role
in determining the temperature scale of the phase coherence.

\begin{figure}[tbp]
\begin{center}
\includegraphics[width=3in] {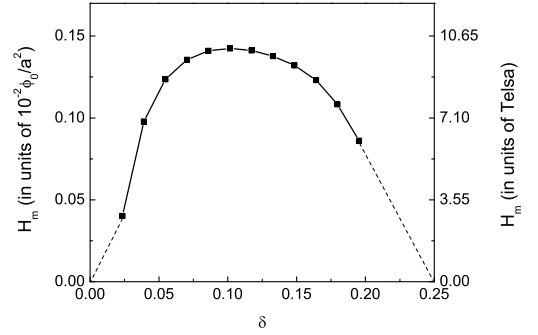}
\end{center}
\caption{Doping dependence of the ``melting'' magnetic field $H_m$
under low
temperature $T=0.01J$, calculated from selfconsistent equation (\protect\ref%
{selfconsistentTc}) with $t_h=3J$. To the right, the label of $H_m$
in the units of Tesla ($\phi_0/a^2=7100.9{\rm T}$) is used. }
\label{Bm}
\end{figure}

Finally, the number of spinon-vortices can be significantly
induced by the magnetic field via the Zeeman coupling as discussed
in Sec. III, which in turn leads to an enhancement of the low
energy spin fluctuations as indicated by the low-temperature Curie
behavior shown in Fig. \ref{chis}. Then it is expected that
$T_{c}$ get quickly reduced by such an enhancement of
the low-lying spin fluctuations via the magnetic-field dependence of $\chi _{0}$%
. Such a suppression of $T_{c}$ by magnetic field is naturally
included in the self-consistent equation (\ref{selfconsistentTc}).
The critical (``melting'') magnetic field $H_{m}$ defined by
$T_{c}(H_{m})=0$ marks a quantum KT transition, where the vortices
melt and proliferate at $T=0$ due to quantum fluctuations. $H_m$ is
plotted as a function of $\delta $ in Fig. \ref{Bm}, with
$T_c(H_m)=0.01\mathrm{J}$ in the numerical calculation.

\begin{figure}[tbp]
\begin{center}
\includegraphics[width=3.2in] {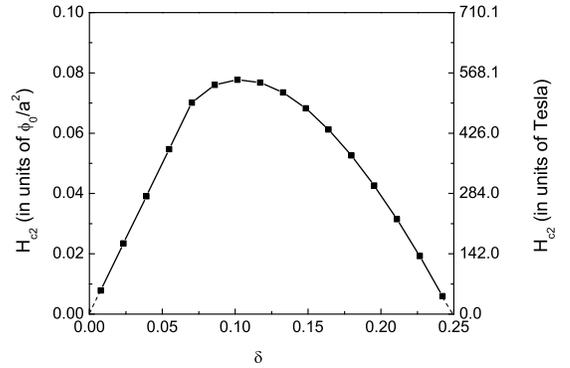}
\end{center}
\caption{Doping dependence of $H_{c2}$ decided by Eq.
(\ref{nv=delta}), calculated for $t_h=3J$ at $T=0.003J$. }
\label{plotHc2}
\end{figure}

\subsection{Upper Phase Boundary}

The upper phase boundary of the SVP is defined by the temperature
and magnetic field scales $T_v$ and $H_{c2}$, respectively, at which
the holon condensation disappears and the effective theory
(\ref{eff}) breaks down. As has been discussed in the
Ginzburg-Landau description in Ref.\onlinecite{weng2006}, such a
phase boundary can be estimated by the core touching of spinon
vortices, which completely destroy the short range phase coherence
of the holon condensate. For completeness, in the following we
provide the estimation of $T_v$ and $H_{c2}$ in the effective theory
(\ref{eff}), which is qualitatively consistent with the
Ginzburg-Landau result but quantitatively different.

The core touching condition can be simply written as\cite{weng2006}
\begin{equation}
n_{v}=\delta  \label{nv=delta}
\end{equation}%
which means the number of vortices is comparable to that of holons.
Since we are now focusing on the upper phase boundary where the
spinon-vortices proliferate, the phase fluctuations discussed in
Sec. IV A is no longer important. Thus we can use the mean-field
equation (\ref{nv}) of Sec. III C to determine the number of
spinon-vortices. The temperature scale $T_{v}$ and magnetic field scale $%
H_{c2} $ of the upper phase boundary are numerically obtained as
shown in Fig. \ref{plotTvHc2} and \ref{plotHc2}, respectively. The
value of $T_{v}$ is reasonable as compared to the
experiment, whearas $H_{c2}$ is about one order of magnitude larger \cite%
{wang2002}, possibly due to the omission of the holon amplitude
fluctuations in the dual theory (\ref{eff}), which are crucial near
the upper phase boundary.

\section{Conclusion and discussions}

In conclusion, we have constructed an effective field theory
description of the spontaneous vortex phase, based on the mutual
Chern-Simons theory of the doped Mott insulator. By introducing a
dual transformation in the holon Lagrangian, the spontaneous vortex
phase, which corresponds to a holon condensate without the
superconducting phase coherence, is described as a spinon-vortex
liquid, in which strong superconducting fluctuations are determined
by the quantum spinon-vortex dynamics. Consequently, both the
residual diamagnetism and spin paramagnetism can be calculated from
the same spin free energy under mean-field approximation, and the
entropy of each vortex can be estimated in this theory, which are
compared with the results of Nernst experiments. The key difference
between the SVP described in the mutual Chern-Simons theory and the
other proposals of vortex liquid phase is the quantum and spinful
nature of vortices, which leads to a closed relation between charge
and spin properties. As shown in Sec. III C, the magneto-resistance
under transverse magnetic field ($H\parallel ab$ plane) is of
comparable size as that under the perpendicular field $H_{\perp }$,
which shows that even the in-plane magnetic-field can change the
vortex number due to the spin Zeeman effect. Another prediction from
the spinon-vortex picture is the existence of spin Hall effect in
the SVP, which suggests that the transverse vortex flow under
external electric field $E_{x}$ and magnetic field $B_{z}^e$ carries
a spin current, as has been discussed in Ref. \onlinecite{kou2005b}.

When considering the physics in the neighborhood of
superconducting phase transition, fluctuations beyond the
mean-field approximation should be included, which leads to a KT
type critical theory with coupling constants
decided by the spinon-vortex correlation functions. The transition temperature $%
T_c$ calculated from this critical theory is shown to be comparable
to the experiments. The magnetic field $H_m$ needed to kill
superconductivity can be calculated in a similar way. The upper
phase boundary $T_v$ and $H_{c2}$ are crossover temperature and
magnetic field scales where the vortex number becomes comparable to
the holon number, and the short range phase coherence of holons is
finally destroyed. In the global phase diagram of mutual
Chern-Simons theory, the SVP is a wide fluctuation region on top of
the superconductivity dome, both of which are embedded in the upper
pseudogap phase with short-range antiferromagnetic
correlations.\cite{gu2005}

There are also several issues that are not included in the present
theory: i) the amplitude fluctuation of holon is not included, so
that the suppression of holon superfluidity under strong magnetic
field is underestimated, which leads to an overestimation of
high-field diamagnetism and $H_{c2}$. ii) the fermionic
quasiparticle is described as a bound state of holon and spinon,
which is well-defined at low energy, long wavelength regime when
spinon excitation is gapped. But in the SVP the spinon is unconfined
and the quasiparticle will decay into spinon and holon, just like
the usual RPA collective mode merging into the particle-hole
continuum. Consequently, the contribution of quasiparticle is not
important in the SVP, although it is important for some low energy
features in the superconducting phase.

In the present work, the onset doping of superconductivity is
$\delta=0$, which is a consequence of ignoring quantum fluctuations
of the holon density. Future work is needed to derive a more
accurate critical theory for the superconducting phase transition---
both the quantum one in the doping axis and the classical one in the
temperature axis.

{\bf Acknowledgement.} We acknowledge helpful discussions with Z. C.
Gu, S. P. Kou, P. A. Lee, L. Li, V. N. Muthukumar, T. K. Ng, N. P.
Ong and Y. Y. Wang. This work is supported by the NSFC grant No.
10374058 and No. 90403016.

\bibliography{hightc}

\begin{thebibliography}{24}
\expandafter\ifx\csname natexlab\endcsname\relax\def\natexlab#1{#1}\fi
\expandafter\ifx\csname bibnamefont\endcsname\relax
  \def\bibnamefont#1{#1}\fi
\expandafter\ifx\csname bibfnamefont\endcsname\relax
  \def\bibfnamefont#1{#1}\fi
\expandafter\ifx\csname citenamefont\endcsname\relax
  \def\citenamefont#1{#1}\fi
\expandafter\ifx\csname url\endcsname\relax
  \def\url#1{\texttt{#1}}\fi
\expandafter\ifx\csname urlprefix\endcsname\relax\def\urlprefix{URL }\fi
\providecommand{\bibinfo}[2]{#2}
\providecommand{\eprint}[2][]{\url{#2}}

\bibitem[{\citenamefont{Bednorz and M$\ddot{\rm u}$ller}(1986)}]{bednorz1986}
\bibinfo{author}{\bibfnamefont{J.~G.} \bibnamefont{Bednorz}} \bibnamefont{and}
  \bibinfo{author}{\bibfnamefont{K.~A.} \bibnamefont{M$\ddot{\rm u}$ller}},
  \bibinfo{journal}{Z. Phys. B} \textbf{\bibinfo{volume}{64}},
  \bibinfo{pages}{189} (\bibinfo{year}{1986}).

\bibitem[{\citenamefont{Kou et~al.}(2005{\natexlab{a}})\citenamefont{Kou, Qi,
  and Weng}}]{kou2005}
\bibinfo{author}{\bibfnamefont{S.-P.} \bibnamefont{Kou}},
  \bibinfo{author}{\bibfnamefont{X.-L.} \bibnamefont{Qi}}, \bibnamefont{and}
  \bibinfo{author}{\bibfnamefont{Z.-Y.} \bibnamefont{Weng}},
  \bibinfo{journal}{Phys. Rev. B} \textbf{\bibinfo{volume}{71}},
  \bibinfo{pages}{235102} (\bibinfo{year}{2005}{\natexlab{a}}).

\bibitem[{\citenamefont{Weng et~al.}(1998)\citenamefont{Weng, Sheng, and
  Ting}}]{weng1998}
\bibinfo{author}{\bibfnamefont{Z.~Y.} \bibnamefont{Weng}},
  \bibinfo{author}{\bibfnamefont{D.~N.} \bibnamefont{Sheng}}, \bibnamefont{and}
  \bibinfo{author}{\bibfnamefont{C.~S.} \bibnamefont{Ting}},
  \bibinfo{journal}{Phys. Rev. Lett.} \textbf{\bibinfo{volume}{80}},
  \bibinfo{pages}{5401} (\bibinfo{year}{1998}).

\bibitem[{\citenamefont{Weng et~al.}(1997)\citenamefont{Weng, Sheng, Chen, and
  Ting}}]{weng1997}
\bibinfo{author}{\bibfnamefont{Z.~Y.} \bibnamefont{Weng}},
  \bibinfo{author}{\bibfnamefont{D.~N.} \bibnamefont{Sheng}},
  \bibinfo{author}{\bibfnamefont{Y.-C.} \bibnamefont{Chen}}, \bibnamefont{and}
  \bibinfo{author}{\bibfnamefont{C.~S.} \bibnamefont{Ting}},
  \bibinfo{journal}{Phys. Rev. B} \textbf{\bibinfo{volume}{55}},
  \bibinfo{pages}{3894} (\bibinfo{year}{1997}).

\bibitem[{\citenamefont{Gu and Weng}(2005)}]{gu2005}
\bibinfo{author}{\bibfnamefont{Z.-C.} \bibnamefont{Gu}} \bibnamefont{and}
  \bibinfo{author}{\bibfnamefont{Z.-Y.} \bibnamefont{Weng}},
  \bibinfo{journal}{Phys. Rev. B} \textbf{\bibinfo{volume}{72}},
  \bibinfo{pages}{104520} (\bibinfo{year}{2005}).

\bibitem[{\citenamefont{Weng and Muthukumar}(2002)}]{weng2002}
\bibinfo{author}{\bibfnamefont{Z.~Y.} \bibnamefont{Weng}} \bibnamefont{and}
  \bibinfo{author}{\bibfnamefont{V.~N.} \bibnamefont{Muthukumar}},
  \bibinfo{journal}{Phys. Rev. B} \textbf{\bibinfo{volume}{66}},
  \bibinfo{pages}{094509} (\bibinfo{year}{2002}).

\bibitem[{\citenamefont{Weng and Qi}()}]{weng2006}
\bibinfo{author}{\bibfnamefont{Z.-Y.} \bibnamefont{Weng}} \bibnamefont{and}
  \bibinfo{author}{\bibfnamefont{X.-L.} \bibnamefont{Qi}},
  \bibinfo{howpublished}{cond-mat/0603097, to appear on PRB}.

\bibitem[{\citenamefont{Fisher and Lee}(1989)}]{fisher1989}
\bibinfo{author}{\bibfnamefont{M.~P.~A.} \bibnamefont{Fisher}}
  \bibnamefont{and} \bibinfo{author}{\bibfnamefont{D.~H.} \bibnamefont{Lee}},
  \bibinfo{journal}{Phys. Rev. B} \textbf{\bibinfo{volume}{39}},
  \bibinfo{pages}{2756} (\bibinfo{year}{1989}).

\bibitem[{\citenamefont{Chen and Weng}(2005)}]{chen2005}
\bibinfo{author}{\bibfnamefont{W.~Q.} \bibnamefont{Chen}} \bibnamefont{and}
  \bibinfo{author}{\bibfnamefont{Z.~Y.} \bibnamefont{Weng}},
  \bibinfo{journal}{Phys. Rev. B} \textbf{\bibinfo{volume}{71}},
  \bibinfo{pages}{134516} (\bibinfo{year}{2005}).

\bibitem[{\citenamefont{Wang et~al.}(2005)\citenamefont{Wang, Li, Naughton, Gu,
  Uchida, and Ong}}]{wang2005}
\bibinfo{author}{\bibfnamefont{Y.}~\bibnamefont{Wang}},
  \bibinfo{author}{\bibfnamefont{L.}~\bibnamefont{Li}},
  \bibinfo{author}{\bibfnamefont{M.~J.} \bibnamefont{Naughton}},
  \bibinfo{author}{\bibfnamefont{G.~D.} \bibnamefont{Gu}},
  \bibinfo{author}{\bibfnamefont{S.}~\bibnamefont{Uchida}}, \bibnamefont{and}
  \bibinfo{author}{\bibfnamefont{N.~P.} \bibnamefont{Ong}},
  \bibinfo{journal}{Phys. Rev. Lett.} \textbf{\bibinfo{volume}{95}},
  \bibinfo{pages}{247002} (\bibinfo{year}{2005}).

\bibitem[{\citenamefont{Mitrovic et~al.}(2003)\citenamefont{Mitrovic, Sigmund,
  Halperin, Reyes, Kuhns, and Moulton}}]{mitrovic2003}
\bibinfo{author}{\bibfnamefont{V.~F.} \bibnamefont{Mitrovic}},
  \bibinfo{author}{\bibfnamefont{E.~E.} \bibnamefont{Sigmund}},
  \bibinfo{author}{\bibfnamefont{W.~P.} \bibnamefont{Halperin}},
  \bibinfo{author}{\bibfnamefont{A.~P.} \bibnamefont{Reyes}},
  \bibinfo{author}{\bibfnamefont{P.}~\bibnamefont{Kuhns}}, \bibnamefont{and}
  \bibinfo{author}{\bibfnamefont{W.~G.} \bibnamefont{Moulton}},
  \bibinfo{journal}{Phys. Rev. B} \textbf{\bibinfo{volume}{67}},
  \bibinfo{pages}{220503} (\bibinfo{year}{2003}).

\bibitem[{\citenamefont{Kakuyanagi et~al.}(2003)\citenamefont{Kakuyanagi,
  Kumagai, Matsuda, and Hasegawa}}]{kakuyanagi2003}
\bibinfo{author}{\bibfnamefont{K.}~\bibnamefont{Kakuyanagi}},
  \bibinfo{author}{\bibfnamefont{K.}~\bibnamefont{Kumagai}},
  \bibinfo{author}{\bibfnamefont{Y.}~\bibnamefont{Matsuda}}, \bibnamefont{and}
  \bibinfo{author}{\bibfnamefont{M.}~\bibnamefont{Hasegawa}},
  \bibinfo{journal}{Phys. Rev. Lett.} \textbf{\bibinfo{volume}{90}},
  \bibinfo{pages}{197003} (\bibinfo{year}{2003}).

\bibitem[{\citenamefont{Harris et~al.}(1995)\citenamefont{Harris, Yan, Matl,
  Ong, Anderson, Kimura, and Kitazawa}}]{harris1995}
\bibinfo{author}{\bibfnamefont{J.~M.} \bibnamefont{Harris}},
  \bibinfo{author}{\bibfnamefont{Y.~F.} \bibnamefont{Yan}},
  \bibinfo{author}{\bibfnamefont{P.}~\bibnamefont{Matl}},
  \bibinfo{author}{\bibfnamefont{N.~P.} \bibnamefont{Ong}},
  \bibinfo{author}{\bibfnamefont{P.~W.} \bibnamefont{Anderson}},
  \bibinfo{author}{\bibfnamefont{T.}~\bibnamefont{Kimura}}, \bibnamefont{and}
  \bibinfo{author}{\bibfnamefont{K.}~\bibnamefont{Kitazawa}},
  \bibinfo{journal}{Phys. Rev. Lett.} \textbf{\bibinfo{volume}{75}},
  \bibinfo{pages}{1391} (\bibinfo{year}{1995}).

\bibitem[{\citenamefont{Ando and Segawa}(2002)}]{ando2002}
\bibinfo{author}{\bibfnamefont{Y.}~\bibnamefont{Ando}} \bibnamefont{and}
  \bibinfo{author}{\bibfnamefont{K.}~\bibnamefont{Segawa}},
  \bibinfo{journal}{Phys. Rev. Lett.} \textbf{\bibinfo{volume}{88}},
  \bibinfo{pages}{167005} (\bibinfo{year}{2002}).

\bibitem[{\citenamefont{C.~Capan et~al.}(2002)\citenamefont{C.~Capan, A.~G.
  M.~Jansen, and Flouquet}}]{capan2002}
\bibinfo{author}{\bibfnamefont{J.~H.} \bibnamefont{C.~Capan},
  \bibfnamefont{K.~Behnia}}, \bibinfo{author}{\bibfnamefont{C.~M. C.~M.}
  \bibnamefont{A.~G. M.~Jansen}, \bibfnamefont{W.~Lang}}, \bibnamefont{and}
  \bibinfo{author}{\bibfnamefont{J.}~\bibnamefont{Flouquet}},
  \bibinfo{journal}{Phys. Rev. Lett.} \textbf{\bibinfo{volume}{88}},
  \bibinfo{pages}{056601} (\bibinfo{year}{2002}).

\bibitem[{\citenamefont{Wang et~al.}(2002)\citenamefont{Wang, Ong, Xu,
  Kakeshita, Uchida, Bonn, Liang, and Hardy}}]{wang2002}
\bibinfo{author}{\bibfnamefont{Y.}~\bibnamefont{Wang}},
  \bibinfo{author}{\bibfnamefont{N.~P.} \bibnamefont{Ong}},
  \bibinfo{author}{\bibfnamefont{Z.~A.} \bibnamefont{Xu}},
  \bibinfo{author}{\bibfnamefont{T.}~\bibnamefont{Kakeshita}},
  \bibinfo{author}{\bibfnamefont{S.}~\bibnamefont{Uchida}},
  \bibinfo{author}{\bibfnamefont{D.~A.} \bibnamefont{Bonn}},
  \bibinfo{author}{\bibfnamefont{R.}~\bibnamefont{Liang}}, \bibnamefont{and}
  \bibinfo{author}{\bibfnamefont{W.~N.} \bibnamefont{Hardy}},
  \bibinfo{journal}{Phys. Rev. Lett.} \textbf{\bibinfo{volume}{88}},
  \bibinfo{pages}{257003} (\bibinfo{year}{2002}).

\bibitem[{\citenamefont{Xu et~al.}(2000)\citenamefont{Xu, Ong, Wang, Kageshita,
  and Uchida}}]{xu2000}
\bibinfo{author}{\bibfnamefont{Z.~A.} \bibnamefont{Xu}},
  \bibinfo{author}{\bibfnamefont{N.~P.} \bibnamefont{Ong}},
  \bibinfo{author}{\bibfnamefont{Y.}~\bibnamefont{Wang}},
  \bibinfo{author}{\bibfnamefont{T.}~\bibnamefont{Kageshita}},
  \bibnamefont{and} \bibinfo{author}{\bibfnamefont{S.}~\bibnamefont{Uchida}},
  \bibinfo{journal}{Nature} \textbf{\bibinfo{volume}{406}},
  \bibinfo{pages}{486} (\bibinfo{year}{2000}).

\bibitem[{\citenamefont{Wang et~al.}(2003)\citenamefont{Wang, Ono, Onose, Gu,
  Ando, Tokura, S.Uchida, and Ong}}]{wang2003}
\bibinfo{author}{\bibfnamefont{Y.}~\bibnamefont{Wang}},
  \bibinfo{author}{\bibfnamefont{S.}~\bibnamefont{Ono}},
  \bibinfo{author}{\bibfnamefont{Y.}~\bibnamefont{Onose}},
  \bibinfo{author}{\bibfnamefont{G.}~\bibnamefont{Gu}},
  \bibinfo{author}{\bibfnamefont{Y.}~\bibnamefont{Ando}},
  \bibinfo{author}{\bibfnamefont{Y.}~\bibnamefont{Tokura}},
  \bibinfo{author}{\bibnamefont{S.Uchida}}, \bibnamefont{and}
  \bibinfo{author}{\bibfnamefont{N.~P.} \bibnamefont{Ong}},
  \bibinfo{journal}{Science} \textbf{\bibinfo{volume}{299}},
  \bibinfo{pages}{86} (\bibinfo{year}{2003}).

\bibitem[{\citenamefont{Wang et~al.}(2006)\citenamefont{Wang, Li, and
  Ong}}]{wang2006}
\bibinfo{author}{\bibfnamefont{Y.}~\bibnamefont{Wang}},
  \bibinfo{author}{\bibfnamefont{L.}~\bibnamefont{Li}}, \bibnamefont{and}
  \bibinfo{author}{\bibfnamefont{N.~P.} \bibnamefont{Ong}},
  \bibinfo{journal}{Phys. Rev. B} \textbf{\bibinfo{volume}{73}},
  \bibinfo{pages}{024510} (\bibinfo{year}{2006}).

\bibitem[{\citenamefont{Harshman and A.~P.~Mills}(1992)}]{harshman1992}
\bibinfo{author}{\bibfnamefont{D.~R.} \bibnamefont{Harshman}} \bibnamefont{and}
  \bibinfo{author}{\bibfnamefont{J.}~\bibnamefont{A.~P.~Mills}},
  \bibinfo{journal}{Phys. Rev. B} \textbf{\bibinfo{volume}{45}},
  \bibinfo{pages}{10684} (\bibinfo{year}{1992}).

\bibitem[{\citenamefont{Kogut}(1979)}]{kogut1979}
\bibinfo{author}{\bibfnamefont{J.~B.} \bibnamefont{Kogut}},
  \bibinfo{journal}{Rev. Mod. Phys.} \textbf{\bibinfo{volume}{51}},
  \bibinfo{pages}{659} (\bibinfo{year}{1979}).

\bibitem[{\citenamefont{Kamal et~al.}(2002)\citenamefont{Kamal, Bonn,
  Goldenfeld, Hirschfeld, Liang, and Hardy}}]{kamal1994}
\bibinfo{author}{\bibfnamefont{S.}~\bibnamefont{Kamal}},
  \bibinfo{author}{\bibfnamefont{D.~A.} \bibnamefont{Bonn}},
  \bibinfo{author}{\bibfnamefont{N.}~\bibnamefont{Goldenfeld}},
  \bibinfo{author}{\bibfnamefont{P.~J.} \bibnamefont{Hirschfeld}},
  \bibinfo{author}{\bibfnamefont{R.}~\bibnamefont{Liang}}, \bibnamefont{and}
  \bibinfo{author}{\bibfnamefont{W.~N.} \bibnamefont{Hardy}},
  \bibinfo{journal}{Phys. Rev. Lett.} \textbf{\bibinfo{volume}{73}},
  \bibinfo{pages}{1845} (\bibinfo{year}{2002}).

\bibitem[{\citenamefont{Osborn et~al.}(2003)\citenamefont{Osborn, Harlingen,
  Aji, Goldenfeld, Oh, and Eckstein}}]{osborn2003}
\bibinfo{author}{\bibfnamefont{K.~D.} \bibnamefont{Osborn}},
  \bibinfo{author}{\bibfnamefont{D.~J.~V.} \bibnamefont{Harlingen}},
  \bibinfo{author}{\bibfnamefont{V.}~\bibnamefont{Aji}},
  \bibinfo{author}{\bibfnamefont{N.}~\bibnamefont{Goldenfeld}},
  \bibinfo{author}{\bibfnamefont{S.}~\bibnamefont{Oh}}, \bibnamefont{and}
  \bibinfo{author}{\bibfnamefont{J.~N.} \bibnamefont{Eckstein}},
  \bibinfo{journal}{Phys. Rev. B} \textbf{\bibinfo{volume}{68}},
  \bibinfo{pages}{144516} (\bibinfo{year}{2003}).

\bibitem[{\citenamefont{Kou et~al.}(2005{\natexlab{b}})\citenamefont{Kou, Qi,
  and Weng}}]{kou2005b}
\bibinfo{author}{\bibfnamefont{S.-P.} \bibnamefont{Kou}},
  \bibinfo{author}{\bibfnamefont{X.-L.} \bibnamefont{Qi}}, \bibnamefont{and}
  \bibinfo{author}{\bibfnamefont{Z.-Y.} \bibnamefont{Weng}},
  \bibinfo{journal}{Phys. Rev. B} \textbf{\bibinfo{volume}{72}},
  \bibinfo{pages}{165114} (\bibinfo{year}{2005}{\natexlab{b}}).

\end{thebibliography}

\end{document}